# Developing a model of evacuation after an earthquake in Lebanon

Hong Van Truong, Elise Beck, Julie Dugdale, and Carole Adam

*Abstract*— **This article describes the development of an agent-based model (AMEL, Agent-based Model for Earthquake evacuation in Lebanon) that aims at simulating the movement of pedestrians shortly after an earthquake. The GAMA platform was chosen to implement the model. AMEL is applied to a real case study, a district of the city of Beirut, Lebanon, which potentially could be stricken by a M7 earthquake. The objective of the model is to reproduce real life mobility behaviours that have been gathered through a survey in Beirut and to test different future scenarios, which may help the local authorities to target information campaigns.**

## I. INTRODUCTION

Every year earthquakes cause many victims. Although the survival of people is largely related to the resilience of buildings, the way that people behave after an earthquake also influences the number of fatalities (Rojo, Beck, Lutoff and Schoeneich, 2011), especially for low magnitude earthquakes. The rareness of strong earthquakes and the impossibility of forecasting them prevent us from assessing the effect that behaviours may have on survival rates. In this context, computer simulation is an appropriate and powerful tool that helps us to assess different scenarios, thus helping local authorities to develop their risk management policies and information campaigns.

The objective of the AMEL model (Agent-based Model for Earthquake evacuation in Lebanon) is to simulate pedestrians' mobility shortly after an earthquake. Although many victims in buildings are generally killed during the main shock, the mobility of any survivors can greatly affect their chances of surviving any aftershocks. We want to understand whether a better knowledge of the safety procedures and the location of safe places in a city could decrease people's vulnerability. Firstly of all, the model aims at reproducing the behaviours adopted by individuals. Secondly, the simulation tests some optimistic or pessimistic fictive scenarios, in order to evaluate the impact of individual behaviours on the number of fatalities or on the number of people in exposed and dangerous situations. This work is applied to the real case of a district of Beirut, Lebanon. This article presents a work in progress and focuses on explaining the model and how it was developed. Section 2 presents the overall approach and gives the state of the art, whilst section 3 describes the methodology used for developing AMEL. The model is described in section 4 and our first attempts at validation are detailed in section 5. Section 6 concludes the paper and presents future work.

## II. APPROACH AND STATE OF THE ART

Agent-based social simulation (ABSS) is a branch of distributed artificial intelligence and multi-agent systems whose focus is on understanding, modelling and simulating social behaviours. Rather than purely focusing on cognition, this approach recognises the social complexity of a society and looks at how social phenomena, such as cooperation, emerge through human behaviours.

The use of agents in modelling human systems has several advantages over other approaches (Bonabeau, 2002). Firstly, agent based systems are able to capture emergent phenomena that are so representative of complex adaptive systems. Secondly, they provide a natural description of a system, which as Bonabeau notes makes the agent based approach much closer to reality. Finally, they are flexible, allowing us to study social systems at different levels of abstraction by varying the complexity of our agents or by aggregating agents into subgroups.

Hundreds of agent-based social simulators have been developed. These have been used for predicting future situations, as training tools, for developing and formalizing theories, or for testing new technological designs or new ways of organizing work (Gilbert and Troitzsch, 1999). Here we focus on the use of a simulator for understanding and predicting human behaviour when faced with an earthquake, depending on their level of vulnerability.

Using agent based simulation as a way to assess emergency situations has become increasingly popular in recent years. The RoboCup Rescue Agents Simulation project has attracted many researchers who are interested in using the specially developed platform to develop intelligent agents that undertake the role of Police Forces, Fire Brigades, and Ambulance Teams in a disaster response scenario (Skinner and Ramchurn, 2010). Likewise, other works, such as the REScUE research project (Hawe, Coates, Wilson and Crouch, 2011) and EQ-Rescue (Fiedrich, 2006) have been developed to evaluate different rescue plans, or to optimize resource allocation during an emergency. These works, like many of this kind of simulation, focus specifically on the response and rescue aspects, with the main accent being on modelling rescue agents, on coordination between different teams, or on the allocation of resources. Our work differs in this respect since we do not look at response activities, but just the behavior of the potential victims.

Pelechano and Badler present a model of building evacuation to study the influence of inter-agent communication and the effect of training (Pelechano and

H. V. Truong is with the Institut de la Francophonie pour l'Informatique, Hanoi, Vietnam. (e-mail: hongvantruongiph@gmail.com).

E. Beck is with the University of Grenoble-Alpes. France (e-mail: Elise.Beck@ujf-grenoble.fr).

J. Dugdale is with the University of Grenoble-Alpes, France and affiliated to the University of Agder, Norway. (e-mail: Julie.Dugdale@imag.fr).

C. Adam is with the University of Grenoble-Alpes, France (e-mail: Carole.Adam@imag.fr).

Badler, 2006). The movements of agents in rooms are modelled by acceleration equations. The authors show that the evacuation may be made more efficient by introducing communication between agents to share their knowledge of blocked routes. Similarly the evacuation is also more efficient if there are a small number of leaders (trained staff) in the crowd that the other dependent agents group around and follow. Although we have designed AMEL to take into account leader/follower behaviours, our work is still in progress and therefore we have not yet experimented with this aspect. Concerning inter-agent communication, we hope to build on this work by considering neighbourhood evacuation.

Nguyen et al. present a hybrid model of the pedestrian flow on road networks, applied to the evacuation of Nhatrang (Vietnam) in case of a tsunami (Nguyen, Zucker, Nguyen, 2011). This model combines micro (agent-based, fine-grained but slow) and macro (equation-based, fast) models of pedestrians' movement, in order to improve the efficiency (speed) of simulations involving a large number of agents. This hybrid model was shown to be more efficient than a micro-model, and of better quality than a macro-model.

## III. METHODOLOGY

A controversy in ABSS concerns the fundamental question of how to develop useful models of real-life social situations. Broadly there are two schools of thought. One follows the KISS (Keep It Simple, Stupid) philosophy where the aim is to develop simplistic models and where much of the real world detail has been abstracted away. Although there are obvious benefits, e.g. in terms of ease of constructing the models, the approach has been widely criticised. The arguments can be reduced to the idea that models that are too simple only address simplistic problems that are not representative of the real world. The other extreme is a KIDS (Keep It Descriptive Stupid) approach (Edmonds and Moss, 2004) where the model is constructed by taking into account the widest possible range of evidence, including anecdotal accounts and expert opinion. Although we obtain a much truer representation of reality it may be very difficult to obtain the data to build the model, implementation is more complicated, and validation of the model and simulator are problematic. The approach adopted in this work falls in between these two extremes and follows that proposed by Rosaria Conte: "Keep it Simple as Suitable" (Conte, 2000). Here models are abstract enough to achieve an adequate level of generality, but no less complex than what is required by the purpose of the simulation. Whilst the original KISS, KIDS, and reformulated KISS approaches provide advice on designing models it is very general and somewhat vague, and the approaches lack a complete modelling method. In response, several methods and modelling techniques have been proposed, for example GAIA (Wooldridge, Jennings and Kinny, 2000), VOWELS (Demazeau, 1995), CoMoMAS (Glaser, 1996), MMTS (Kinny, Geogeff and Rao, 1996), and Unified Approach (Sabas, Delisle and Badr, 1996). These all provide the standard framework for modelling the agent dimension, some taking into account the deliberative behaviour of agents. However, they are largely intended for developing general multi-agent systems and are not specifically focused on modelling the social elements that are required in ABSS. Furthermore, they fail to provide a structure for analysing human agents in the design phase and for validating the model with respect to the observed human behaviour. Some methods have been developed that focus particular on agent based simulation. Notably the ODD protocol by Grimm and his colleagues aimed to standardising descriptions of models to aid understanding and ensure reproducibility (Grimm, Berger, Bastiansen, Eliassen, Ginot, Giske, Goss-Custard, Grand, Heinz, Huse, Huth, Jepsen, Jørgensen, Mooij, Müller, Pe'er, Piou, Railsback, Robbins, Robbins, Rossmanith, Rüger, Strand, Souissi, Stillman, Vabø, Visser and DeAngelis DL. 2006; Grimm, Berger, DeAngelis, Polhill, Giske and Railsback, 2010). Although very useful, the protocol only concerns model description and does not suggest how the model itself may be developed.

The methodology adopted in this work is shown in figure 1 and is adapted from that of Edmonds (Edmonds, 2000). However it has several important differences. Firstly, it focuses on analysing human behaviour in the real world situation through the use of extensive field studies. This provides a solid corpus of empirical data through video recordings and observations, etc. Secondly, it puts validation at the heart of the process ensuring that the results of the simulator can be more readily trusted (Dugdale, Bellamine-Ben Saoud, Pavard, Pallamin, 2001). Finally, it reinforces iteration; this allows us to revisit previous steps such as undertaking additional targeted field studies and refining the model and code as necessary.

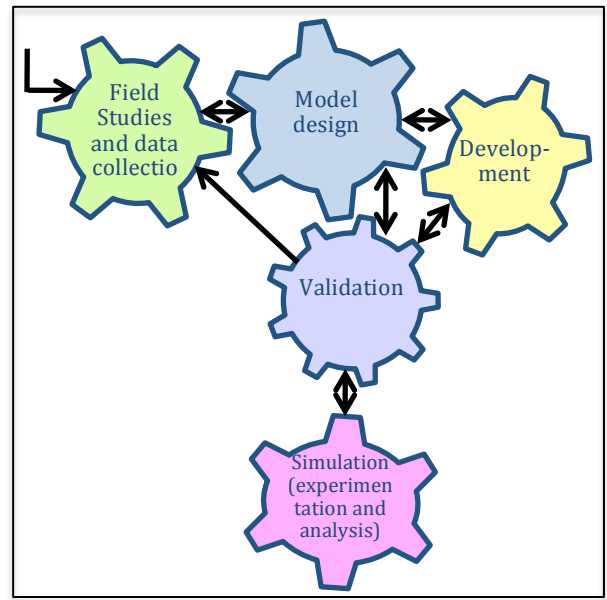

Figure 1. Methodology adopted for AMEL

Following figure 1, the first step covers performing detailed field studies of the real situation in order to assess the human behaviours and their underlying motivations. The second step, Model design, involves developing the formal model, for example by using UML, of what has been obtained through the analysis of field studies data. Validation has been put at the heart of the methodology and relies heavily on the data obtained via field studies in the analysis phase. The final step, Simulation, covers experimentation.

The bi-directional arrows ensure that iteration plays a major role. The above methodology was first described in 1999 (Dugdale, Pavard and Soubie, 1999; Dugdale, Pavard and Soubie 2000) and has been used for designing and developing agent-based simulators in several works over the years, for example in (Bellamine-Ben Saoud, Ben Mena, Dugdale, Pavard and Ahmed, 2006; Kashif, Binh, Dugdale and Ploix, 2001; Kasif, Ploix, Dugdale and Binh, 2013).

IV. FIELD STUDIES AND DATA COLLECTION

The studied district is characterized by a high density of buildings that have a differing number of storeys, date and construction material. Several types of data were collected using a multidisciplinary approach, involving the disciplines of geography, psychology and geotechnics. Concerning the geographic aspect, the buildings, streets, and green spaces correspond to spatial data represented by polygons and polylines. These objects correspond to geographic layers and were processed through geographic information systems. The data first comes from a national reference database (such as the French Geographic Institute database) and was corrected and updated using high spatial resolution satellite imaging. Concerning demographic data, the last Lebanon population census was carried out in 1932. Therefore, the population was estimated at 3.8 persons per apartment.

Because of the high heterogeneity of the building types, which leads to high variations of physical vulnerability to earthquakes, each of the 357 buildings was documented through a field survey that aimed at characterizing its vulnerability. For example, the survey form included information such as number of storeys, year of construction, construction material. Some geotechnical data was collected with material samples for different types of construction. This data was then computed with the FEMA's method called HAZUS (Kircher, Whitman and Holmes, 2006) in order to estimate the damage rate for each building in case of an M7 earthquake on the Yammouneh fault.

Finally, the model requires social data concerning the behaviours of individuals. This data was collected through a field sociological survey that interviewed 88 persons of the studied district (Beck, Colbeau-Justin, Cartier and Saikali, 2011). The questionnaire focused on several subjects including knowledge about earthquakes, risk perception, earthquake experience and associated behaviours, etc. The statistical analysis of the survey allowed us to define several categories of behaviours. Some extra behaviours, that were not observed in the survey but have been reported in similar cases, were also considered and implemented into the model (leader-follower behaviour, for example).

V. MODEL DESIGN

As a first step, the purpose of the simulation is to reproduce, based on the survey information, the mobility of people and their behaviours following an earthquake in Lebanon. In an earthquake, natural and artificial obstacles (e.g. escarpments and stairways), and obstacles induced by the earthquake itself, can injure people and constrain their movement towards safe zones. The simulator will provide information concerning the physical and human damage incurred (the number of damaged buildings, blocked streets, number of victims, number of people in dangerous zones, evacuation time, etc.).

Once we have reproduced mobility behaviours that align to the survey information, as a second step, we experiment with difference scenarios by changing the ways that people behave and analysing the results. We want to see what happens (damages, number of fatalities and exposure to danger) if people act in different ways. These results can serve to inform people about how to improve their behaviours, for example by forming the basis of an information campaign.

*A. Entities and their attributes*

The model consists of six entities: Human, Street, Building, Green space, Quake, and Obstacle (figure 2). Their attributes are divided into three types: position and visualization (location, colour, etc.), entity characteristics (e.g. Human entity has attributes about age, sex, profession, etc., and the Building entity has attributes about height, capacity, etc.). Finally there are special attributes concerning the entity's behaviours (e.g. target to reach, street_knowledge, etc. of the Human entity).

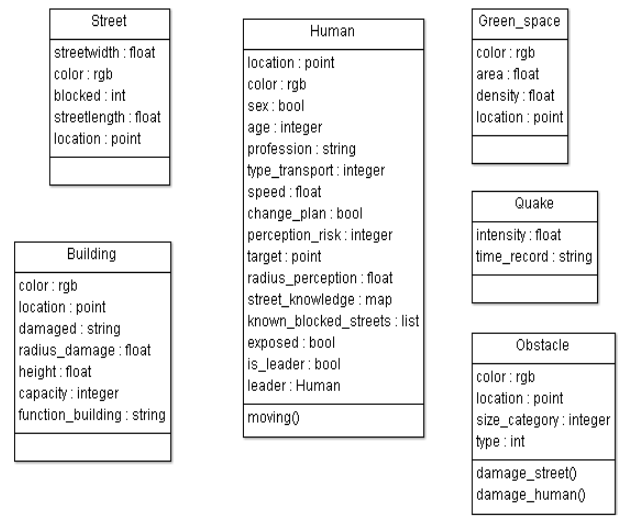

Figure 2. Class diagram

*B. Human agent*

Since the Human agent is the most important entity we focus on describing this agent in detail. Human agents represent people in the district. After an earthquake, agents can decide to move, on roads that are not blocked, or stay at their current location. They can become informal leaders who know the safe areas and can guide other agents to their targets. Agents can also become followers by searching for a leader in their perception zones and then following that agent. If a follower cannot find a leader, it wanders around.

Human agents can observe obstacles and other agents inside its perception radius (attribute radius_perception). After meeting a big obstacle, the Human agent can act in different ways. If it is a leader, it will choose another way to reach target. If it is a wanderer, it will choose another target to go to. Each Agent update their street knowledge, remembering which streets are blocked so that they may be avoided in future.

## C. Obstacle agent

Obstacle agents can affect the behaviour of Human agents. In this work, we focus on the obstacles induced by an earthquake itself, particularly by the buildings damaged by the earthquake. Based on the damage level of buildings, we can group obstacles into three levels: big, medium and small. The big obstacles destroy all the streets and people in those zones, so no agent can pass these obstacles. In the zone of medium and small obstacles, people are exposed to the danger (attribute exposed); this means that the agent is in the danger zone and there is a high risk of being injured. Medium obstacles can constrain the movement of agents in their zone, making them move more slowly.

## D. Other agents

The Quake agent has attributes for the intensity and time of the earthquake. Building agents have attributes for damage level of the building. Street agents are considered as a weighted graph where each street is an edge and each intersection is a vertex. Each unaffected street has a small weight (e.g. the length of street), but blocked streets have a much higher weight than normal streets (e.g. a billion). By using this weighted graph approach each agent has their own representation (beliefs) about the state of the streets.

The spatial scale in our model is in meters and each simulation cycle lasts one second. The simulation begins after all human agents have decided upon their target, and ends when all the leaders have reached their targets.

## E. Agent behaviours

Human behaviours:

How people behave in reality is very complex. In our model, we just focus on the mobility of people and do not consider more complex behaviours. From the survey, we synthesize 6 behaviours:

1. Move without changing activity. This covers people that had intended to move and do not change their activity as a result of the earthquake (e.g. people who are on their way to work)
2. Stay in place without changing activity. This covers people who remain where they are and continue doing the same activity (e.g. those who are at home and do not have employment).
3. Change activity in order to go to a safe place. This covers people who intentionally move to a safe place just after an earthquake.
4. Change activity in order to go to an unsafe place. This covers peoples who intentionally move to an unsafe place (e.g. someone who goes to a relative's home even though it may be unsafe).
5. Change activity and stay in place. This covers people who had planned to go somewhere, but because of the earthquake they remain where they are.
6. Imitate others. This addresses leader-follower behaviours.

The simulation begins when all of the agents have determined where to go. Thus, from a mobility point of view, behaviours 1, 3 and 4 are equivalent (the agent moves), and behaviours 2 and 5 are also equivalent (the agent does not move).

Human behaviours are synthesized into three groups: movement, obstacle perception and imitation. While movement behaviour concerns the mobility of people, obstacle perception and imitation relate to determining and updating target locations and knowledge of the blocked streets.

- Movement behaviour

The streets are considered as a weighted graph on which the Human agent moves and determines the shortest path to reach its target. Moreover, the street knowledge of agent is considered as list of weights applied for the graph. Each agent has a different knowledge of the streets, so the weights applied for the graph are different according to each agent. Thus the shortest path to the target depends on two factors: the agent's target and the agent's street knowledge (figure 3).

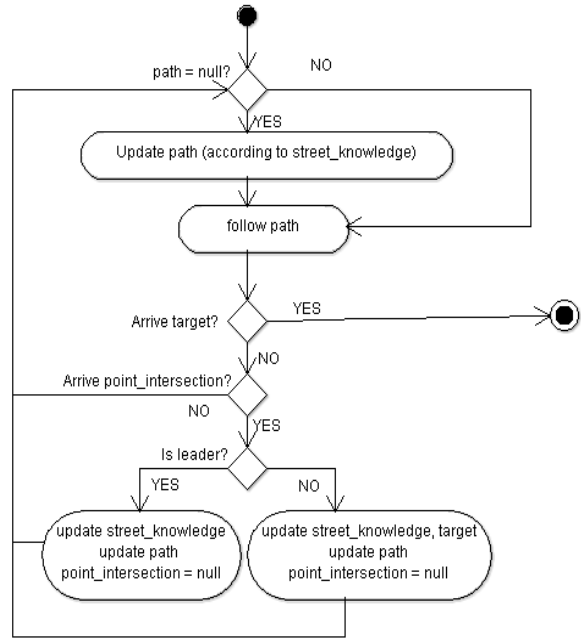

Figure 3. The movement process

- Obstacle perception behaviour

Agents within the zone of medium and small obstacles are considered to be exposed to danger. In addition, medium obstacles constrain the movement of agents, making them move along their route more slowly.

Agents can observe obstacles in their zone of perception and when they encounter a big obstacle they cannot proceed. However, they remember blocked streets so that they may be avoided in the future when finding another path to the target. The difference between the behaviour of leaders and followers is that if they meet an obstacle, leaders will find other ways to reach their target, while followers will find another target to go to. In our model, we consider 5 blocked streets configurations (Fig. 4).

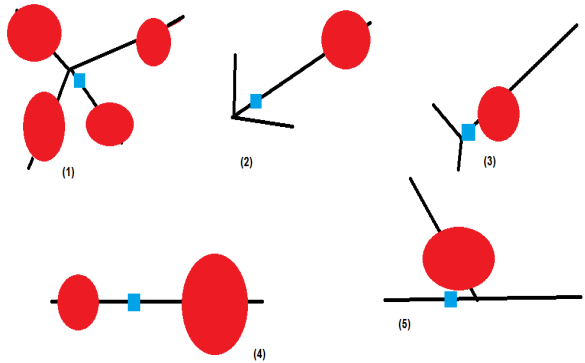

Figure 4. Blocked street configurations (with damages)

In figure 4 the red circle is an obstacle and a blue square is the Human agent. In (1) the Agent is in the area with no exit. In (2) there is a long street section before the agent moves towards the obstacle. In (3) there is a short street section before the agent moves towards the obstacle. In (4) the Agent is in the street with two obstacles on both sides. In (5) the Agent sees an obstacle on another street.

The simplest method to implement this behaviour is to update an agent's knowledge of blocked streets and target path when the agent encounters an obstacle. However, this leads to several problems.

First, with configuration (2), the agent is still in the blocked street, it must go back to choose another street, but the street section to go back to is longer than the rest of the street, so the agent often chooses to enter into the obstacle. To overcome this problem we make all agents update their street knowledge and go back to the previous intersection after meeting an obstacle (Figure 5). After that, the weights of blocked streets are updated, leader agents build other paths to reach their targets, and wandering agents choose other targets and build paths to go. If an agent encounters an obstacle as in the configuration (5), figure 4, then it will normally update the weights.

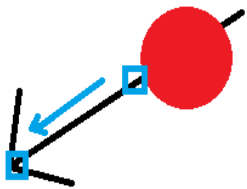

Figure 5. Agent moves to the previous intersection to update its street knowledge

Secondly, with configurations (1) and (4), the agent has no choice but to pass over obstacles to get out. In order to fix this problem we add a parameter concerning the maximum number of times that the agent encounters an obstacle (2-3 times). After this, the agent stops and stays where it is. This feature relates to the psychological aspect of people in the evacuation. If a person tries to find a way to its target but repeatedly doesn't have success it will eventually give up. Figure 6 shows the obstacle perception behaviour.

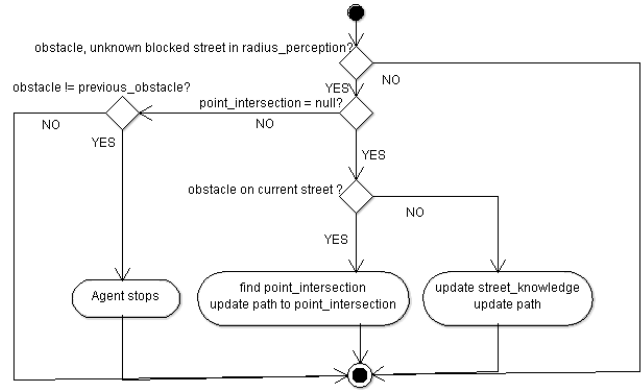

Figure 6. Obstacle perception process

- Imitation behaviour

For the followers and wanderers, at the beginning, they search randomly around them for a leader. If they find a leader, then they become followers. They update their targets to the position of leader and update paths to follow it. Figure 7 shows the process of imitation.

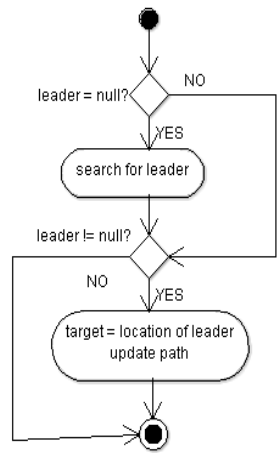

Figure 7. Imitation process for followers and wanderers

For the leaders, every 10 cycles of the simulation, they perceive their followers and adjust their speeds to the slowest follower (figure 8).

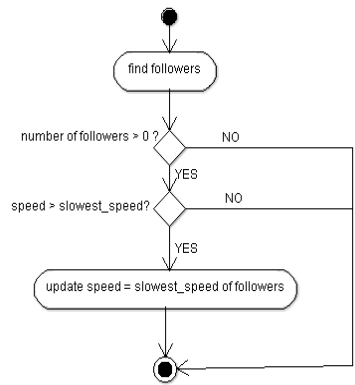

Figure 8. Speed adjustment process for leader

- Obstacle behaviours

The Obstacle agent has two behaviours: street damage and human damage. With street damage, each agent finds all the streets around it that it overlaps and updates the "blocked" attribute of that Street agent to "true". With human damage, each Obstacle agent finds all of the Human agents with which it overlaps and destroys them.

## VI. Development

The AMEL model has been implemented on the GAMA platform (Grignard, Taillandier, Gaudou, Huynh and Drogoul, 2013); GAMA is an open-source generic agent-based modelling and simulation platform. It provides an intuitive modelling language with high-level primitives to define agents and their environment. GAMA includes a powerful Integrated Development Environment to help non-computer scientists to develop complex models with powerful features in terms of Geographical Information Systems (GIS) integration and high-level tools (e.g. decision-making or clustering algorithms). In addition, both the language and the software have been designed to allow the development of big models with a huge number of agents (with various architectures from reflex-based to belief-desire-intention architectures). GAMA also allows modellers to manage various kinds of complex environments such as square, hexagonal or irregular grids, networks, or continuous environments linked to GIS data.

GAMA enables a hierarchical and dynamic organization of agents. The platform is also easily extensible to add new features to models (e.g. to give the possibility of integrating equation-based (ODE) models into agents) or new agent architectures or features. GAMA has been successfully used to develop various large-scale applications that share the need for a tight integration and management of huge GIS data, and for strong interactions between a complex environment and the agents. It was used for instance in the MAELIA platform for simulating water management problems (Gaudou, Sibertin-Blanc, Thérond, Amblard, Arcangeli, Balestrat, Charron-Moirez, Gondet, Hong, Mayor, Panzoli, Sauvage, Sanchez-Perez, Taillandier, Nguyen, Vavasseur and Mazzega, 2013).

Figure 9 shows a screenshot of the simulation. The main display represents the district with the streets in black, green space in green, and the buildings in yellow. Human agents are represented by small circles of different colours: hot colours (red, pink) for agents who stay where they are; cold colours (blue, purple) for agents who move; and the colour cyan for imitating agents. We also created other displays: charts of the different behaviours; charts about the number of victims, people exposed, leaders and followers, etc; and a chart about the total exposition time. On the right, we can see the parameters of the simulator; these concern the location of the human agents and their behaviours, as well as some global simulator parameters. By varying the value of these parameters we can create different scenarios which may help to see the effect of the different information campaigns. For example, if we want to see the effect on the number of victims if an earthquake occurs during the night, we can set the value for the "people in buildings" parameter (probability of people who are in the buildings) equal to 100 (percent), indicating that at night everyone is inside a building.

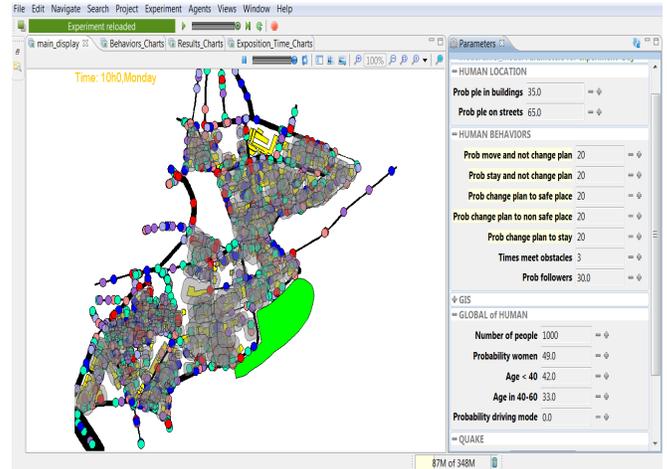

Figure 9. Screenshot of the simulation

## VII. Validation

Once the model has been implemented, it must be validated. This is current work in progress. Every newly implemented "brick" of the model is tested individually and all the associated scenarios are evaluated in order to understand the impact of the new brick. As we developed the simulator incrementally it eventually becomes increasingly complex. The challenge is to understand the impact of any new changes on the global model. For example, once human exposure to hazards is implemented, we have to verify that the fatalities figures are realistic. For these kinds of validations we ask the experts who collected the data and those that have real life experience of the consequences of earthquakes. This process of validation involves all other steps of the methodology and is central to the development process. In practice we have found this to be a highly iterative step where we frequently request validation from experts when changing our model and after implementation. This validation constitutes a major step that should not be underestimated.

## VIII. First results

Although we are currently in the validation stage, we provide some first results so that the reader may see the type of results that may be achieved with the simulator. The results concern two scenarios: 1. The survey scenario, this is the real situation where most people stay where they are and do not go to the safe place; 2. An optimistic scenario where everyone goes to a safe place (e.g. green space, schools, outside of the city, or even just into the streets).

We can see in the graphs below that the number of exposed people in the optimistic scenario is less than that in the survey scenario. Also the number of 'others' (i.e. those who are not exposed and injured, in other words they are in the safe places) and those who go to outside of the city (considered as a safe place) in the optimistic scenario is more than in the survey scenario.

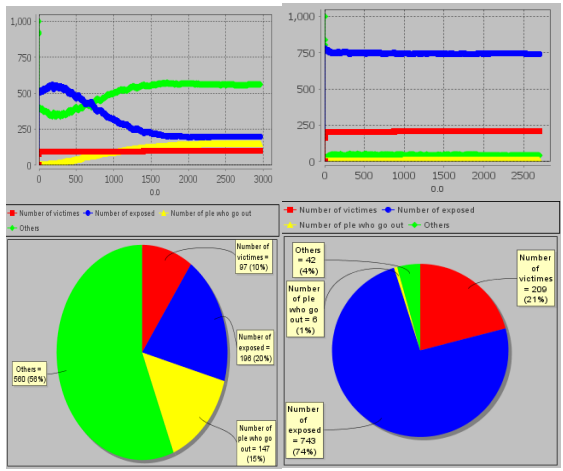

*Figure 10. Fatality and exposure graphs. On the left, the survey scenario. On the right an optimistic scenario*

Concerning the total exposure time, we can see that in the current situation (survey scenario) there are more people who are exposed for a long time, whereas in the optimistic scenario there are more people who are exposed for the short time.

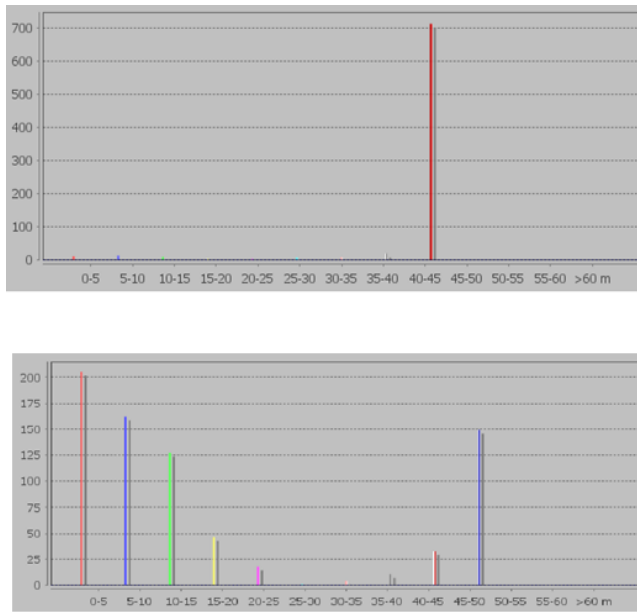

*Figure 11. Number of agents exposed. On top, the survey scenario. On the bottom the optimistic scenario. The horizontal axis shows time slot periods and the vertical axis shows the number of exposed people in a corresponding time period.*

## IX. SUMMARY

Lebanon is an earthquake prone area, lying on the Mount Lebanon Thrust Fold and the Yamouneh faults, which have the capacity to generate a 7.5 magnitude earthquake. Although major earthquakes are rare, minor ones occur frequently and there is a deep concern that a large earthquake could destroy the city, as it did in 551 AD. Despite this threat the study conducted as part of this project showed that, possibly because people are constantly exposed to many small earthquakes, a significant number of them do not change their behaviours during an earthquake and continue to do their originally planned activities. Whilst in small earthquakes this is not problematic, the consequences of being unprepared in the case of a large one could be devastating.

We have developed a model and first prototype of a simulator that shows pedestrians' movements following different human behaviours after an earthquake. Using GAMA we were able to incorporate the actual map of the neigbourhood in Beirut, including current streets and actual buildings. This allowed us to develop a realistic simulator with agents moving from known buildings along current streets. One particular feature of how we have designed the model is that each agent has its own personal view of the state of the road infrastructure. This replicates the real situation where people are only aware of the routes open to them after an earthquake as they move around the neigbourhood towards their target. As agents explore their neigbourhood, their knowledge of the streets is incrementally updated.

We are currently in the process of validating the model and simulator. In particular we are addressing the problem of an agent's exposure to danger and how fatalities may be quantified given the different severities of building damage. Validation is a long process since we need to interact with local experts from different domains and the data required is not always available. Nevertheless, despite some difficulties with validation, the results of the simulator are encouraging.

ACKNOWLEDGEMENTS

This study is funded by the French National Research Agency (ANR RISKNAT 2009 LIBRIS). The authors would like to thank the different persons involved in the collection of data: Jocelyne Adjizian-Gérard, Rita Zaarour, Nada Saliba, Pierre-Charles Gérard and the students involved in the survey for the building inventory (Geography Department of the Saint Joseph University, Beirut, Lebanon); Jacques Harb for the damages modelling (Notre Dame University, Beirut, Lebanon); Maud Saikali and Carine Azzam (Saint Joseph University, Beirut, Lebanon), Ludvina Colbeau-Justin and Stéphane Cartier (CNRS-PACTE Laboratory, Grenoble, France) for the risk perception survey. The authors also thank Patrick Taillandier (Rouen University, France), Nicolas Marilleau (Institute for Research and Development, France) and Benoit Gaudou (Toulouse University, France) for their precious help on GAMA. Julie Dugdale would like to acknowledge the support of the University of Agder, to which she is affiliated.

REFERENCES

[1] N. Bellamine-Ben Saoud, T. Ben Mena, J. Dugdale, B. Pavard and B. Ahmed, "Assessing large scale emergency rescue plans: an agent based Approach. Special Issue on Emergency Management Systems. International Journal of Intelligent Control and Systems. Vol. 11, No. 4, Dec. 2006. 260-271.

[2] M. Bertran Rojo, E. Beck, C. Lutoff and Ph. Schoeneich, "Exposition sociale face aux séismes : la mobilité en question. Le cas de Lorca (Espagne) – May 2011". *Géorisques*.

[3] Bonabeau E. "Agent-based modeling: Methods and techniques for simulating human systems". In: *Adaptive Agents, Intelligence, and Emergent Human Organization: Capturing Complexity through Agent-Based Modeling. Proceedings of the National Academy of Sciences (PNAS)* U S A. 2002 May 14; 99(Suppl 3): pp. 7280–7287.

[4] R. Conte, "The necessity of intelligent agents in social simulation". In Ballot, G. and Weisbuch, G. (eds.), *Applications of Simulation to Social Sciences*. Paris: Hermes Science, 2000.

[5] Y. Demazeau, "From interactions to collective behaviour in agent based-systems". In *Proceedings of the First Conference on Cognitive Science*. Saint-Malo, France, 1995.

[6] J. Dugdale, N. Bellamine-Ben Saoud, B. Pavard and N. Pallamin, "Simulation and Emergency Management".In Van de Walle, B., Turoff, M. and Hiltz, R.H. (eds) *Information Systems for Emergency Management*. Series: Advances in Management Information Systems. Sharp, 2001.

[7] J. Dugdale, B. Pavard and J.L. Soubie, "Design Issues in the Simulation of an Emergency Call Centre*"*. In *Proceedings of the 13th European Simulation Multiconference (ESM 99)*. June 1-4, 1999. Warsaw, Poland.

[8] J. Dugdale, B. Pavard and J.L. Soubie, "A Pragmatic Development of a Computer Simulation of an Emergency Call Centre". In Dieng, R. (ed) *Designing Cooperative Systems. Frontiers in Artificial Intelligence and Applications*. IOS Press, 2000.

[9] B.Edmonds, "The Use of Models - making MABS actually work". In. Moss, S. and Davidsson, P. (eds.), *Multi Agent Based Simulation*, 2000. Lecture Notes in Artificial Intelligence, 1979:15-32.

[10] B. Edmonds and S. Moss, "From KISS to KIDS: An "anti-simplistic" modelling approach". *In Proceedings of Multi-Agent Based Simulations (MABS) Conference*, 2004, pp. 130-144.

[11] F. Fiedrich, An HLA-based multiagent system for optimized resource allocation after strong earthquakes. Winter Simulation Conference 2006, 486-492.

[12] B. Gaudou, C. Sibertin-Blanc, O. Thérond, F. Amblard, J.-P. Arcangeli, M. Balestrat, M.-H. Charron-Moirez, E. Gondet, Y. Hong, Th. Louail, E. Mayor, D. Panzoli, S. Sauvage, J.-M. Sanchez-Perez, P. Taillandier, V.B. Nguyen, M. Vavasseur and P. Mazzega, "The MAELIA multi-agent platform for integrated assessment of low-water management issues" (regular paper). In *International Workshop on Multi-Agent-Based Simulation (MABS 2013*), Saint-Paul, MN, USA, 06/05/2013-07/05/2013, 2013 (to appear).

[13] N. Gilbert and K.G. Troitzsch, *Simulation for the Social Scientist*. Open University Press, London, 1999.

[14] N. Glaser, *Contribution to Knowledge Acquisition and Modelling in a Multi-Agent Framework — The CoMoMAS Approach*, PhD Thesis Universite Henry Poincare, Nancy, F, December 1996.

[15] A. Grignard., P. Taillandier, B. Gaudou, N.Q. Huynh, D.-A. Vo and A. Drogoul A. *GAMA v. 1.6: Advancing the art of complex agent-based modeling and simulation.* PRIMA 2013.

[16] Grimm V, Berger U, Bastiansen F, Eliassen S, Ginot V, Giske J, Goss-Custard J, Grand T, Heinz S, Huse G, Huth A, Jepsen JU, Jørgensen C, Mooij WM, Müller B, Pe'er G, Piou C, Railsback SF, Robbins AM, Robbins MM, Rossmanith E, Rüger N, Strand E, Souissi S, Stillman RA, Vabø R, Visser U, DeAngelis DL. 2006. A standard protocol for describing individual-based and agent-based models. *Ecological Modelling* 198:115-126.

[17] Grimm V, Berger U, DeAngelis DL, Polhill G, Giske J, Railsback SF. 2010. The ODD protocol: a review and first update. *Ecological Modelling* 221: 2760-2768

[18] Hawe, G.I., Coates, G., Wilson, D.T. & Crouch, R.S**.** (2011). Design Decisions in the Development of an Agent-Based Simulation for Large-Scale Emergency Response. 8th International Conference on Information Systems for Crisis Response and Management, Lisbon, Portugal.

[19] D. Kinny, M. Georgeff and A. Rao, "A Methodology and Modeling Technique for Systems of BDI agents". In *Proceedings of the Seventh European Workshop on Modelling Autonomous Agents in a Multi-Agent World (MAAMAW)*. LNAI, 1996, Volume 1038, 56-71. Springer-Verlag.

[20] T.N.A. Nguyen, J.-D. ZuckerD., H.D. Nguyen., A. Drogoul. and D.-A. Vo, "A Hybrid Macro-Micro Pedestrians Evacuation Model to Speed Up Simulation in Road Networks", In *Advanced Agent Technology workshop at AAMAS 2011*, Taipei, Taiwan, May 2-6, 2011. Revised selected papers. Pp 371-383. Lecture Notes in Computer Science, volume 7068. Springer Berlin Heidelberg. 2012.

[21] N. Pelechano and N. I. Badler. "Modeling Crowd and Trained Leader Behavior during Building Evacuation", *IEEE Computer Graphics and Applications*, Volume 26, Issue 6, November-December 2006, pages 80-86.

[22] A. Sabas, S. Delisle and M. Badri, "A Comparative Analysis of Multiagent System Development Methodologies: Towards a Unified Approach", In *Proceedings of the Third International Symposium From Agent Theory to Agent Implementation, 16th European Meeting on Cybernetics and Systems Research*, 1996.

[23] Skinner, C. and Ramchurn, S. (2010) The RoboCup Rescue simulation platform, Proceedings of the 9[th] International Conference on Autonomous Agents and Multiagent Systems, pp. 1647-1648, Toronto, Canada.

[24] M. Wooldridge, N.R. Jennings. and D. Kinny, "The GAIA Methodology for Agent-Oriented Analysis and Design", *Journal of Autonomous Agents and Multi-Agent Systems*, 2000, 3(3) 285-312.